\documentclass[letterpaper,twocolumn,english,prb,aps,superbib,tightenlines,floatfix]{revtex4}

\usepackage{graphicx,graphicx,epsfig}
\usepackage{hyperref}

\begin{document}

\title{Solid-state memcapacitor}

\author{J. Martinez$^1$, M. Di Ventra$^2$, Yu. V. Pershin$^{1}$}
\affiliation{$^1$Department of Physics and Astronomy and USC
Nanocenter, University of South Carolina, Columbia, SC, 29208 USA
\\ $^2$Department
of Physics, University of California, San Diego, La Jolla, CA
92093-0319, USA}

\begin{abstract}
We suggest a possible realization of a solid-state memory
capacitive (memcapacitive) system. Our approach relies on the slow
polarization rate of a medium between plates of a regular
capacitor. To achieve this goal, we consider a multi-layer
structure embedded in a capacitor. The multi-layer structure is
formed by metallic layers separated by an insulator so that
non-linear electronic transport (tunneling) between the layers can
occur. The suggested memcapacitor shows hysteretic charge-voltage
and capacitance-voltage curves, and both negative and diverging
capacitance within certain ranges of the field. This proposal can
be easily realized experimentally, and indicates the possibility
of information storage in memcapacitive devices.
\end{abstract}

\maketitle

\section{Introduction}

A recent experimental demonstration of a nanoscale memory-resistor
(memristor for short)~\cite{strukov08} has sparked numerous
investigations in the area of materials and systems that show
history-dependence in their current-voltage
characteristics~\cite{pershin08, pershin09, driscoll09a,
driscoll09b, Waser2007-1, Scott2007-1}. This has also led to the
recent proposal and theoretical investigation of memcapacitors and
meminductors, namely capacitors and inductors whose capacitance
and inductance, respectively, depends on the past states through
which the system has evolved~\cite{diventra09}. Together with the
memristor, the whole class of these memory-circuit elements
promises new and unexplored functionalities in electronics, and
the combination of these memory devices with their ``standard''
counterpart may find application in disparate areas of science and
technology, including the study of neuromorphic circuits to
simulate biological processes~\cite{ourPointofView}. In this
paper, we focus only on memcapacitors.

Various systems are known to exhibit memcapacitive behavior
including vanadium dioxide metamaterials \cite{driscoll09a},
nanoscale capacitors with interface traps or embedded nanocrystals
\cite{Kim2001-1,fleetwood95,su05,lee06}, and elastic capacitors
\cite{Nieminen2002-1,partensky2002-1}. As anticipated in Ref.~\onlinecite{diventra09}, memcapacitive effects may
also accompany memristive effects in nanostructures, since in many of
them the morphology of conducting regions changes in time
\cite{strukov08}. Moreover, memcapacitors can be simulated by
electronic circuits \cite{pershin2009-5}. However, the number of
experimentally identified memcapacitive systems is still very
limited. In addition, it would be of great importance to identify
a mamcapacitive system that can be fabricated relatively easily,
and is flexible enough to cover a wide range of capacitances.

\begin{figure}[b]
\centering
\includegraphics[width=8.5cm]{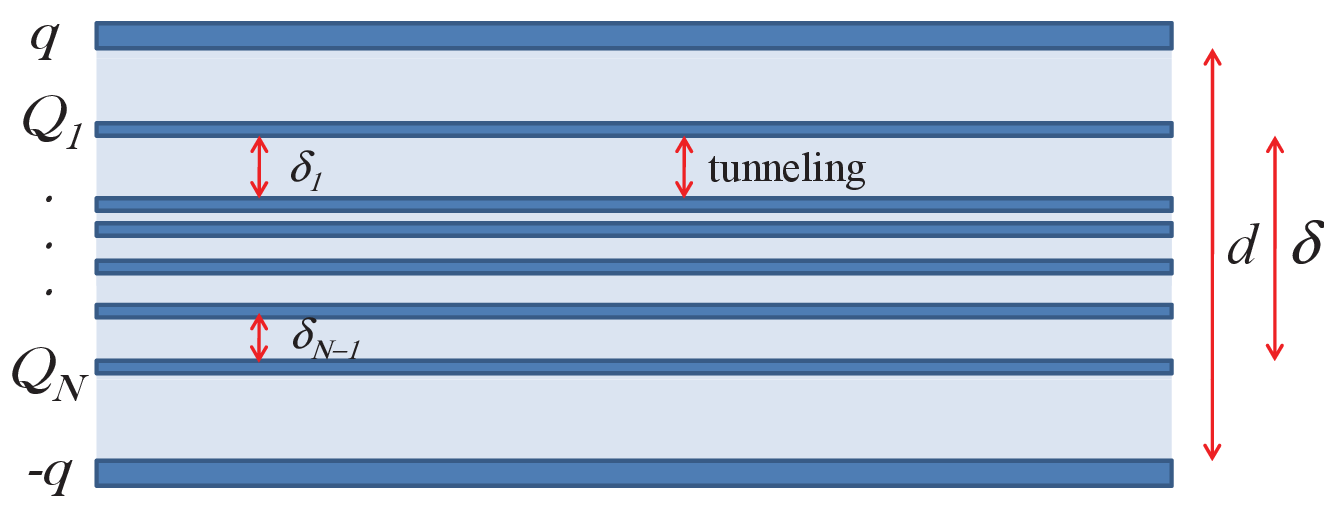}
\caption{General scheme of a solid-state memcapacitor. A
metamaterial medium consisting of $N$ metal layers embedded into
an insulator is inserted between the plates of a ``regular'' capacitor.
It is assumed that the electron transfer between external plates
of the capacitor (with charge $\pm q$) and internal metal layers
(with charges $Q_k$) is negligible. Therefore, the internal charges
$Q_k$ can only be redistributed between the internal layers
creating a medium polarization. \label{fig1}}
\end{figure}

Here, we suggest a possible realization of such a system based on the slow
polarization of a medium between a regular capacitor plates. There are
several physical mechanisms that can potentially provide a slowly
polarizable medium. Examples include tunneling, activated drift of charged
vacancies/impurities, slow ion penetration through a membrane, etc.
In this paper, we will consider the tunneling mechanism and
discuss a solid-state memcapacitive system consisting of metal
layers embedded into a parallel-plate capacitor as schematically shown
in Fig. \ref{fig1}. In this realization, the internal
metal layers, together with the insulator material, form a ``metamaterial''
characterized by a long polarization/depolarization time. The
application of an external voltage to the capacitor leads to a
charge redistribution between the embedded metal layers. The tunneling
current between the layers depends almost exponentially on the
applied voltage. This feature is important for the operation of our suggested device allowing for the
``writing'' of information (in the form
of medium polarization) with high-voltage pulses, and ``holding'' such information when low/zero
voltages are applied. The resulting memcapacitor exhibits not only hysteretic charge-voltage
and capacitance-voltage curves but also both negative and diverging
capacitance within certain ranges of the field. Due to its simplicity and unusual capacitance
features we thus expect
it to find use in both logic and memory applications.

This paper is organized as follows. Sec. \ref{sec1} describes the
theoretical framework used to simulate solid-state memcapacitors. Numerical simulations for the case of ac and pulsed
applied voltages are presented in Sec. \ref{sec2} and \ref{sec3},
respectively. An equivalent circuit model is given in Sec.
\ref{sec4}. Finally, Sec. \ref{sec5} presents our conclusions.

\section{Model} \label{sec1}

Quite generally, a circuit element with memory (memristor,
memcapacitor or meminductor) is defined by the relations~\cite{diventra09}
\begin{eqnarray}
y(t)&=&g\left(x,u,t \right)u(t) \label{Geq1}\\ \dot{x}&=&f\left(
x,u,t\right) \label{Geq2}
\end{eqnarray}
where $u(t)$ and $y(t)$ are any two input and output variables
(i.e., current, charge, voltage, or flux), $g$ is a generalized
response, $x$ is a set of $n$ state variables describing the
internal state of the system, and $f$ is a continuous
$n$-dimensional vector function. In this paper we will be
concerned with only memcapacitors. Therefore, using Eqs.
(\ref{Geq1}, \ref{Geq2}), we define a charge-controlled
memcapacitive system by the equations

\begin{eqnarray}
V_C(t)&=&C^{-1}\left(x,q,t \right)q(t) \label{Ceq1} \\
\dot{x}&=&f\left( x,q,t\right) \label{Ceq2}
\end{eqnarray}
where $q(t)$ is the charge on the capacitor at time $t$, $V_C(t)$
is the applied voltage, and $C$ is the {\it memcapacitance}, which
depends on the state of the system and can vary in time. Below, we
identify the internal state variables of the solid-state
memcapacitor shown in Fig. \ref{fig1} and demonstrate that such
memcapacitor is indeed described by Eqs. (\ref{Ceq1},\ref{Ceq2}).

Let us then consider the system schematically shown in Fig. \ref{fig1} consisting of
$N$ layers of metallic regions separated by an insulating material. This multi-layer system, of thickness $\delta$, is embedded into a standard
parallel-plate capacitor. For convenience, although this is not necessary, we assume that the insulating material in between the
embedded metallic layers is the same as the one of the regular capacitor. We also assume that our
memcapacitor is designed in such a way that there is no electron
exchange between the external plates and those embedded. This can be
achieved by selecting an appropriate separation between the
internal metal layers and the capacitor plates and/or by using an
insulating material between the external capacitor plates
and the embedded ones that generates a higher potential barrier. Therefore, in what follows, we assume that the
tunneling only occurs between the internal metallic layers.
Consequently, the total internal charge is always zero:
$\sum_{k=1}^NQ_k(t)=0$, where $Q_k(t)$ is the charge on the internal metal layers at time $t$.

The operation of the polarization-based memcapacitor relies on the
redistribution of the internal charges $Q_k$ between the embedded
layers caused by the electric field due to the charge $q$ on the
capacitor plates. Polarization of the metamaterial results in an
internal electric field between the layers that is opposite to the
electric field due to the plate charges. Therefore, for a given
amount of plate charge, the internal charges tend to decrease the
plate voltage or, equivalently, to increase the capacitance.

Let us then calculate the internal charge distribution and consequent capacitance.
A charged plane creates a uniform electric field in the direction
perpendicular to the plane with a magnitude $E=\sigma/(2
\varepsilon_0 \varepsilon_r)$, where $\sigma=q/S$ is the surface (of area $S$)
charge density, $\varepsilon_0$ is the vacuum permittivity, and
$\varepsilon_r$ is the relative dielectric constant of the insulating
material. Using this expression, we find that, in the
structure shown in Fig. \ref{fig1}, the external plate voltage is given by,
\begin{eqnarray}
V_C&=&2dE_q+\delta
E_1+\left[\delta-2\delta_1\right]E_2+\left[\delta-2\left(\delta_1+\delta_2\right)\right]E_3+\nonumber\\&+&...+\left[\delta-2\left(\delta_1+...+\delta_{k-1}\right)\right]E_k...-\delta
E_N, \label{eq:VC0}
\end{eqnarray}
where $E_q=q/(2S\varepsilon_0 \varepsilon_r)$ is the electric
field due to the charge $q$ at the external plate and
$E_k=Q_k/(2S\varepsilon_0 \varepsilon_r)$ is the electric field
due to the charge $Q_k$ at the $k$-th embedded metal layer. Eq.
(\ref{eq:VC0}) can also be written as follows
\begin{eqnarray}
V_C&=&\frac{q}{C_0}\left[1+\Delta
\frac{Q_1}{2q}+\left(\Delta-2\Delta_1\right)\frac{Q_2}{2q}+...+\right.\nonumber\\&&\left.+\left(\Delta-2\Delta_{k-1}\right)\frac{Q_k}{2q}...-\Delta\frac{Q_N}{2q}\right],
\label{eq:VC1}
\end{eqnarray}
where $\Delta=\delta/d$, $\Delta_i=\sum_{j=1}^{i}{\delta_j}/d$ for
$i=1,2,...,N-1$, $\Delta_0=0$, and $C_0=\varepsilon_0 \varepsilon_r S/d$ is the capacitance of
the system without internal metal layers. Therefore, the capacitance of the
whole structure is
\begin{equation}
C=\frac{q}{V_C}=\frac{2\,C_0}{2+\sum_{i=1}^{N}\left[\Delta-2\Delta_{i-1}\right]\frac{Q_i}{q}}.
\label{eqC1}
\end{equation}

From this equation it is already clear that, unlike a conventional
capacitor, there may be instants of time in which the denominator
of Eq.~(\ref{eqC1}) is zero while the numerator is finite. This
may happen when the internal metal layers screen completely the
external field, despite the presence of a finite charge $q$ on the
external capacitor plates. At these times one would then expect
{\it diverging} values of capacitance. In addition,
Eq.~(\ref{eqC1}) does not enforce a positive capacitance. Indeed,
as we will show in the examples below, at certain instants of time
the internal metal layers may {\it over-screen} the external
field, resulting in a {\it negative} capacitance. It is
interesting to note that an hysteretic negative and diverging
capacitance has been found also in ionic memcapacitors, namely
nanopore membranes in an ionic solution subject to external
time-dependent perturbations~\cite{krems2009-2}. A negative
capacitance has been experimentally observed in different
solid-state systems \cite{Ershov1998-1,Gommans2005-1,Mora2006-1},
but not accompanied by hysteretic and diverging values of
capacitance.

In order to describe the dynamics of the internal charges $Q_k$,
let us define $V_k=-E_{k,k+1}\delta_k$ as the voltage between $k$
and $k+1$ metal layers. The electric field $E_{k,k+1}$ between two
neighboring layers is obtained by adding the electric fields due to charges
at all layers and at the external metal plates:
\begin{eqnarray}
E_{k,k+1} = -2E_q-E_1-E_2...-E_k+E_{k+1}...+E_N=  \nonumber
\\ =\frac{-2q-\left(Q_1+...+Q_k\right)+\left(Q_{k+1}+...+Q_N\right)}{2S\varepsilon_0
\varepsilon_r}. \;\; \label{el_field}
\end{eqnarray}

The dynamics of the charge at a metal layer $k$ is then determined by the
currents flowing to and from that layer:
\begin{equation}
\frac{dQ_k}{dt}=I_{k-1,k}-I_{k,k+1}, \label{eq:charge_dynamics}
\end{equation}
where $I_{k,k+1}$ is the tunneling electron current flowing from
layer $k$ to layer $k+1$ (note that for the top and bottom layers
($k=1,N$), there is only one term on the right hand side of Eq.
(\ref{eq:charge_dynamics})). By considering the layer charges
$Q_k$ as state variables, we immediately notice that Eq.
(\ref{eq:charge_dynamics}) is similar to Eq. (\ref{Ceq2}). This
demonstrates that the system under consideration is indeed a
charge-controlled memcapacitive system.

\begin{figure}[bt]
\centering
\includegraphics[width=7cm]{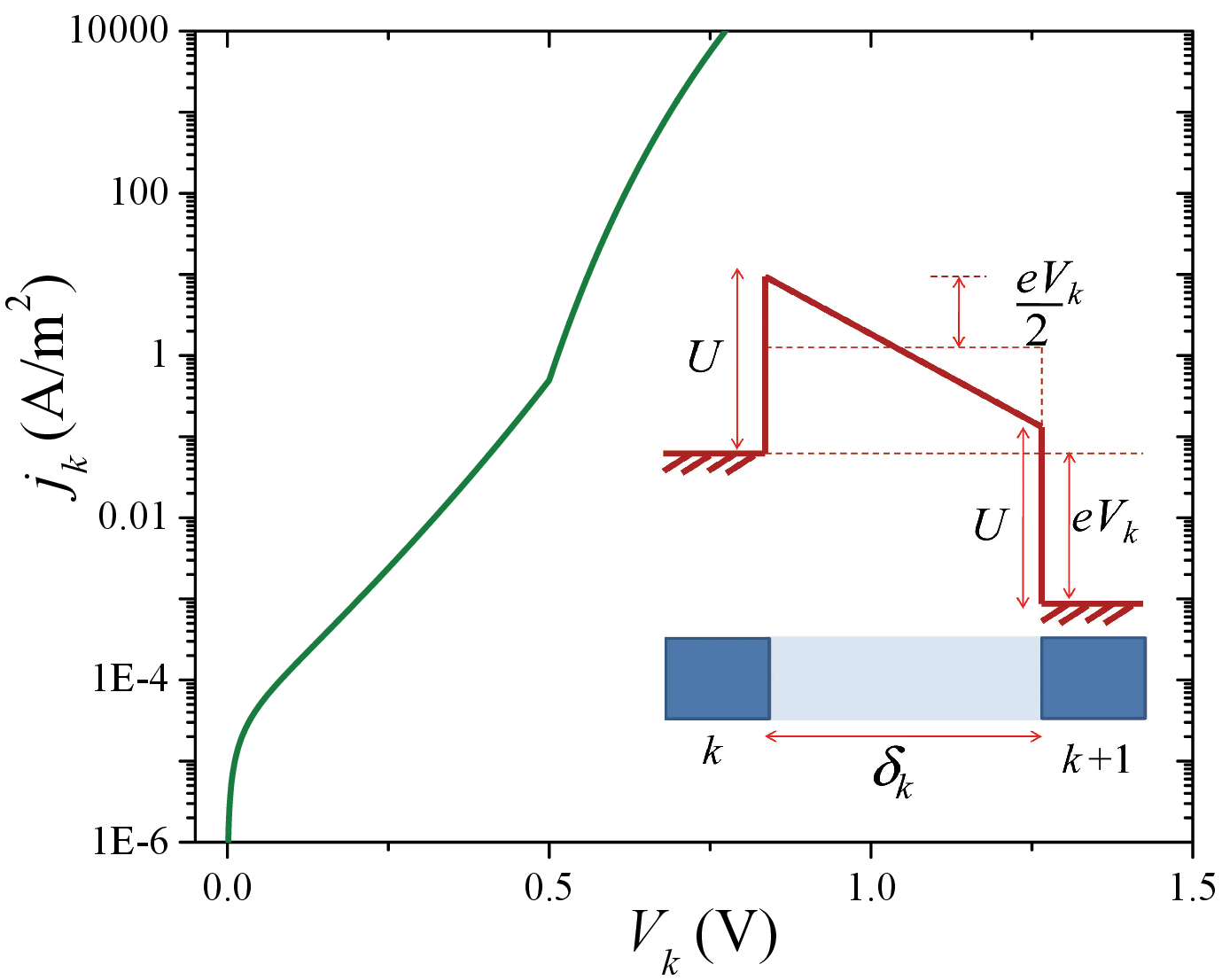}
\caption{Tunneling current density as a function of voltage drop
between two adjacent internal layers calculated using the following values of
parameters: $U=0.5$eV, $\delta_k=5$nm, and $m=m_e$, with $m_e$ the electron mass. Inset: the
energy level scheme. \label{fig2}}
\end{figure}

If $U$ is the potential barrier height between two
adjacent metal layers (see inset of
Fig. \ref{fig2}), the tunneling current induced by the voltage difference $V_k$ between the
two can be calculated using the following expressions \cite{Simmons1963-1}

\begin{figure*}[t]
\centering \includegraphics[width=16cm]{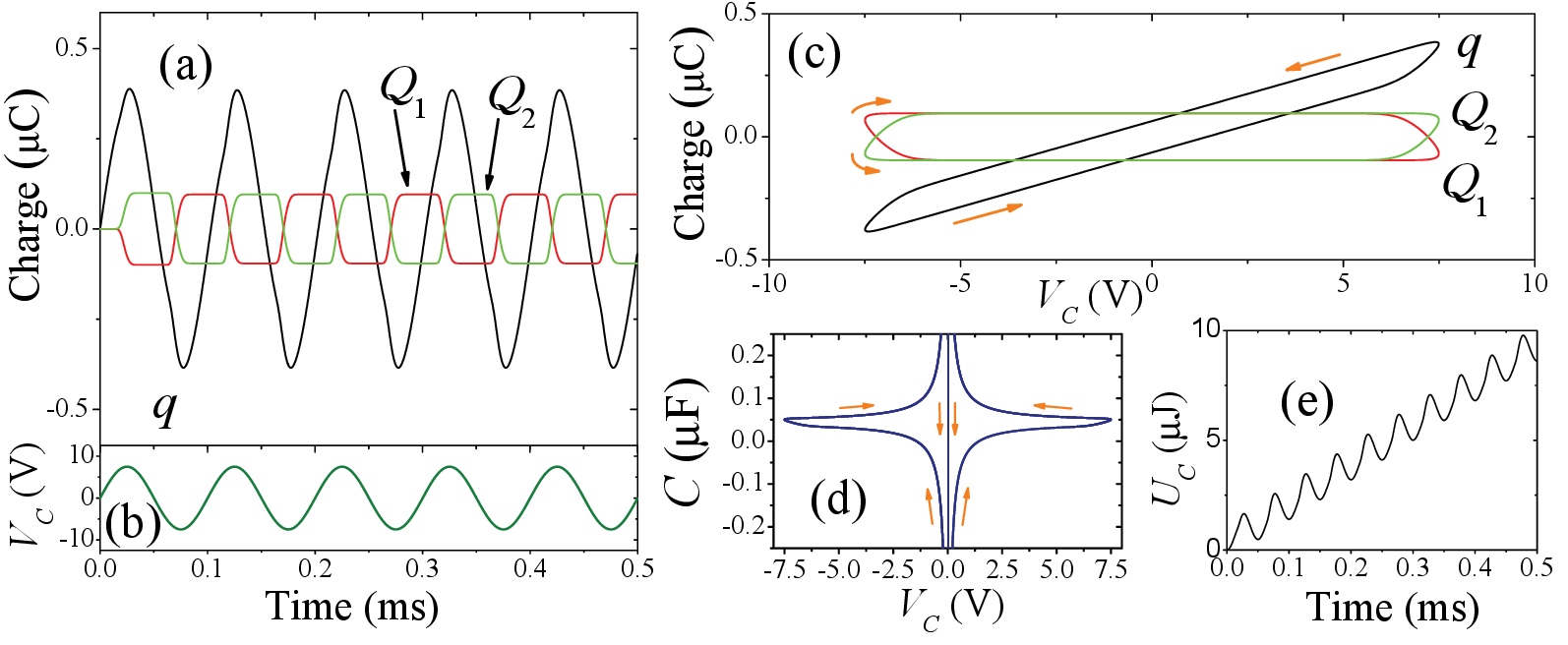}
\caption{Simulation of two-layer memcapacitor with symmetrically
positioned internal layers. (a) The charge on internal metallic
layers and memcapacitor plates as a function of time $t$. (b)
Voltage on memcapacitor, $V_C$, as a function of time $t$.
Charge-voltage (c) and capacitance-voltage (d) plots. (e)
Added/removed energy as a function of time $t$. These plots were
obtained using the parameter values $V_0=7.5$V, $f=10$kHz,
$d=100nm$, $\delta=66.6$nm, $S=10^{-4}$m$^2$, $\varepsilon_r=5$,
$U=0.33$eV, $R=1\Omega$.} \label{fig3}
\end{figure*}

$$ I_{k,k+1}= \frac{S\,e}{2\pi
h\delta_k^2}\left[\left(U-\frac{eV_k}{2}\right)\exp\left[-\frac{4\pi
\delta_k\sqrt{2m}}{h}\sqrt{U-\frac{eV_k}{2}}\right]\right. $$
\begin{equation}
 \left.-\left(U+\frac{eV_k}{2}\right)\exp\left[-\frac{4\pi
\delta_k\sqrt{2m}}{h}\sqrt{U+\frac{eV_k}{2}}\right]\right]
\label{eq:Itun1}
\end{equation}
if $eV_k<U$, and
\begin{eqnarray}
I_{k,k+1}=\frac{S\,e^3\,V_k^2}{4\pi
hU\delta_k^2}\left[\exp\left[-\frac{4\pi
\delta_k\sqrt{m}U^{3/2}}{ehV_k}\right]\right. \nonumber \\ \left.
-\left(1+\frac{2eV_k}{U}\right)\exp\left[-\frac{4\pi
\delta_k\sqrt{m}U^{3/2}}{ehV_k}\sqrt{1+\frac{2eV_k}{U}}\right]\right]
\label{eq:Itun2} \;\;\;\;\;\;\;\;\;\;
\end{eqnarray}
if $eV_k>U$. In the above equations, $h$ is the Planck constant.
The tunneling current density $j_k=I_{k,k+1}/S$ as a function of
voltage drop $V_k$ is depicted in Fig. \ref{fig2} for two adjacent
layers. A strong increase in current with increase of $V_k$ is
clearly observed. This is an important property for the operation
of our suggested memcapacitor.

\section{AC voltage operation} \label{sec2}

In this Section, we present results of numerical simulations obtained
for the case of a sinusoidal voltage $V(t)=V_0 \sin (2 \pi f t)$ of amplitude $V_0$ and
frequency $f$
applied to a memcapacitor $C$ connected in series with a resistor
$R$. Such a circuit is described by the equation
\begin{equation}
V(t) = R\,\frac{dq}{dt}+\frac{q}{C}, \label{eq:circuit}
\end{equation}
where, for $C$, we use Eq. (\ref{eqC1}). Eq. (\ref{eq:circuit}) is
solved numerically together with Eq. (\ref{eq:charge_dynamics})
describing the internal charge dynamics. Below, we present results
of numerical simulations for two different memcapacitor
structures. In our simulations, a small value of $R=1 \Omega$ is
selected so that the voltage drop on the resistor is negligible.

\subsection{2-layer memcapacitor}

The simplest system exhibiting polarization memory is a two-layer
memcapacitor. In Fig. \ref{fig3} we illustrate simulation results
of such a device. Starting at the zero-charge state, the
sinusoidal applied voltage induces electron tunneling between the
internal metal layers resulting in non-zero $Q_1$ and $Q_2$. As it
is shown in Fig. \ref{fig3}(a), positive half-periods of $V$
induce negative $Q_1$ and positive $Q_2$ charges. In turn, these
cause a screening electric field opposite to the electric field of
plate charges.

It is interesting that on the charge-voltage plot (Fig.
\ref{fig3}(c)), $q(V_C)$ forms a non-pinched hysteresis loop. This
is a quite unusual feature of a memory device, especially, in view
of the fact that, at the present time, only memristive devices
exhibiting pinched hysteresis loops have been observed
experimentally. Physically, it is clear that when an internal
polarization in the memcapacitor is present and the plate charge
is zero ($q=0$), the internal polarization creates a non-zero
voltage drop on the device $V_C\neq 0$. Alternatively, this
voltage drop can be compensated by plate charges, but then we get
$V_C=0$ at $q\neq 0$. This explains why the curve does not pass
through the (0,0) point. The corresponding capacitance hysteresis
is shown in Fig. \ref{fig3}(d). As expected, we find both negative
and diverging capacitance values. In the vicinity of $V_C=0$, the
capacitance changes from $+\infty$ to $-\infty$ at both sweep
directions. In Fig \ref{fig3}(d), these abrupt changes appear
as two coinciding vertical lines.

\begin{figure}[tb]
\centering
\includegraphics[width=7cm]{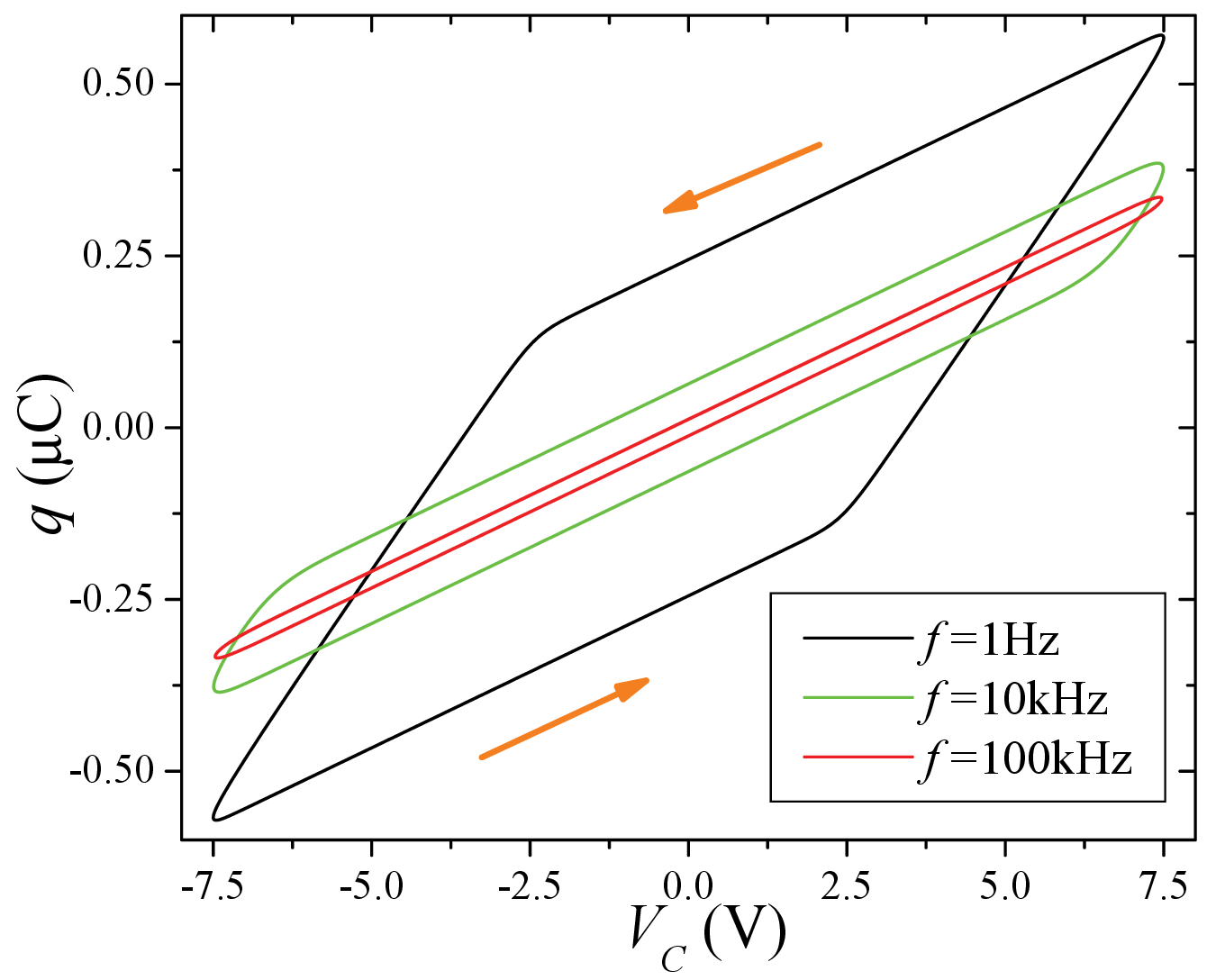}
\caption{Charge-voltage plot at different applied voltage
frequencies $f$. The decrease of the hysteresis at higher
frequencies is a signature of memcapacitors \cite{diventra09}. The
calculation and device parameters are as in Fig. \ref{fig3}.}
\label{fig4}
\end{figure}

\begin{figure*}[!bt]
\centering \includegraphics[width=16cm]{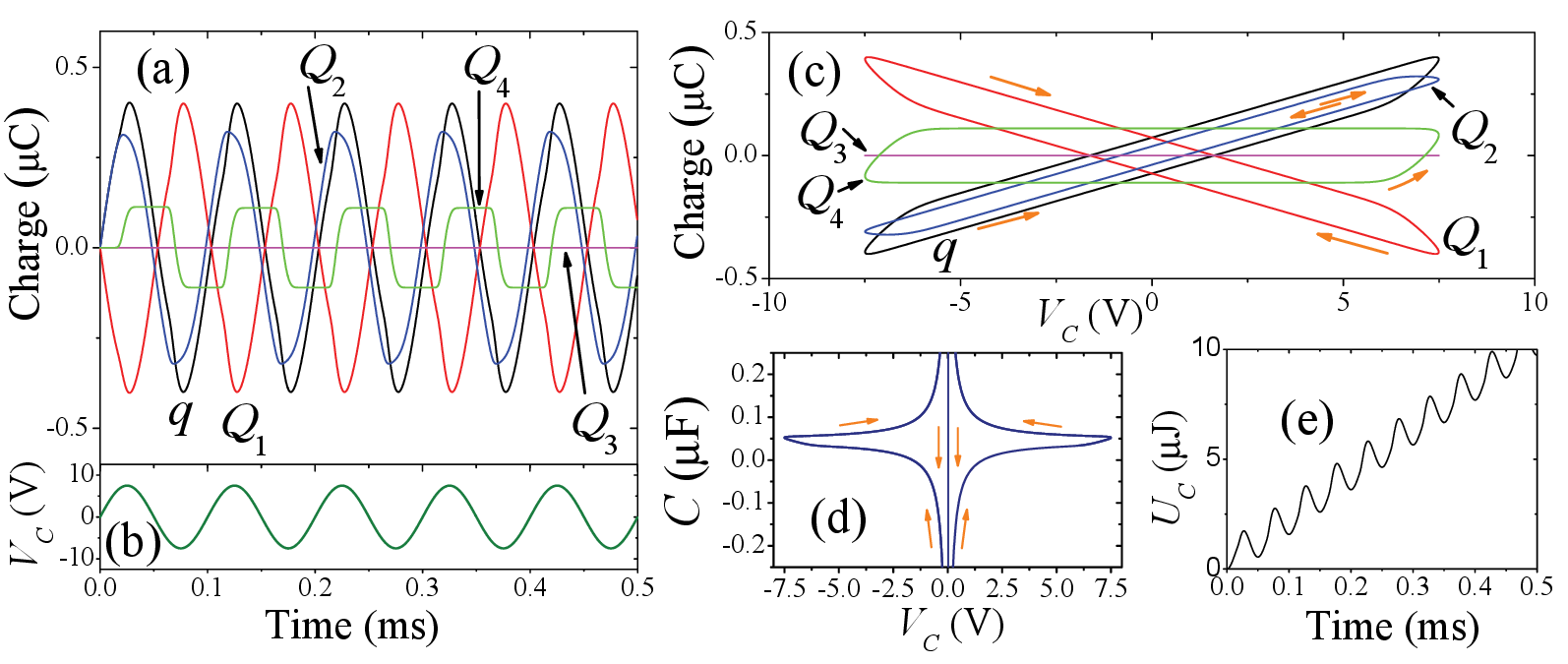}
\caption{Simulation of four-layer memcapacitor. (a) The charge on
internal layers and memcapacitor plates as a function of time $t$.
(b) Voltage on memcapacitor $V_C$ as a function of time $t$.
Charge-voltage (c) and capacitance-voltage (d) plots. (e)
Added/removed energy as a function of time $t$. These plots were
obtained using the same parameter values as in Fig. \ref{fig3}.
The locations of internal layers are defined by parameters
$\delta_1=0.024\delta$, $\delta_2=\delta_3=0.488\delta$.}
\label{fig5}
\end{figure*}

Next, we consider the added/removed energy to/from the capacitor
which is defined as
\begin{equation}
U_C=\int\limits_0^t V_C(\tau )I(\tau )d\tau.
\end{equation}
From Fig. \ref{fig3}(e) it is clear that the present solid-state
memcapacitor operates as a dissipative device since the amount of
added energy is on average larger than the amount of removed
energy (resulting in positive values of $U_C$ at all times). This
occurs because the electron tunneling between internal layers is
accompanied by energy dissipated in thermalization processes due
to the different electrochemical potential energies of metal
layers. In addition, in a real device we cannot exclude the
phenomenon of local heating that would also contribute to
dissipation of energy~\cite{diventra08}.

In fact, the energy dissipation in electron transfer processes or the
usual energy dissipation in dielectric materials described by the
imaginary part of the complex permittivity can be used as an alternative to
our approach. The heat released can drive a phase transition in a
material between the plates, such as transition from crystalline
to amorphous state and vice versa, accompanied by a change in
material's permittivity. In this way, the memcapacitor acquires a
long-term memory based directly on the value of $\varepsilon_r$.
The required material for this purpose can be similar to
chalcogenide glass, which can be ''switched'' between two states,
crystalline or amorphous, with the application of heat, but
possibly at a lower phase transition temperature. The device
operation protocol can then be based in applications of ac-modulated
voltage pulses whose amplitude, duration or frequency can control
the material's state.

Finally, the frequency behavior of the charge-voltage hysteresis loop is
shown in Fig. \ref{fig4}. It is clearly seen that the
hysteresis shrinks at higher frequencies. This is a typical
behavior of memory devices \cite{diventra09} related to the fact
that at high frequencies the internal degrees of freedom of a
memory device do not have enough time to respond to the external perturbation. Similarly, with
increasing frequency, we have observed a decrease in capacitance
hysteresis as well as in the rate of energy dissipation.

\subsection{Multi-layer memcapacitor}

The results obtained for multi-layer memcapacitors are similar to
the two-layer memcapacitor case. The only interesting difference
is that in the case of multi-layer memcapacitors, a richer
internal charge dynamics can be observed. Charge dynamics is
essentially determined by the distance between the layers and the potential energy barrier between adjacent layers. As a prototype of
multi-layer memcapacitor, we consider a four-layer
memcapacitor.

In our particular example two additional internal layers (2-nd and
3-rd) were inserted between the two layers of the two-layer
memcapacitor. The layers' positions are specified in the caption
of Fig. \ref{fig5}: layer 2 is positioned close to layer 1, while
layer 3 is placed in between layers 2 and 4. As a result, the
electron tunneling current can be high between 1-st and 2-nd
layers which is reflected in high amplitudes of $Q_1$ and $Q_2$.
Concerning the charge on the third layer, it is essentially close
to zero. We explain it by a tendency of maximum charge separation
in these types of devices. Fig. \ref{fig5} displays our simulation
results for this case. One can also see that the capacitance
hysteresis and the added/removed energy (Fig. \ref{fig5} (d) and
(e), respectively) are similar to those of the two-layer
memcapacitor as shown in Fig. \ref{fig3}.

\section{Retrieving the memcapacitor state} \label{sec3}

At this point, an important question is how one can read the
memcapacitor state. Here, we demonstrate that the most simple way
to do it is by using a single voltage pulse. Small amplitude or
short voltage pulses are not suitable for this purpose as they do
not change the charge distribution on the internal layers
considerably. Therefore, the measured value of capacitance, using
small amplitude or short pulses, would always be equal to $C_0$,
the capacitance in the absence of the internal metal layers. It is
thus important to ensure that the applied pulse has a large enough
amplitude and is sufficiently long. Consequently, the device state
will be altered by the pulse. However, since the initial state is
read by this measurement, this state can be, in principle,
restored.

\begin{figure}[tb]
\centering
\includegraphics[width=7.5cm]{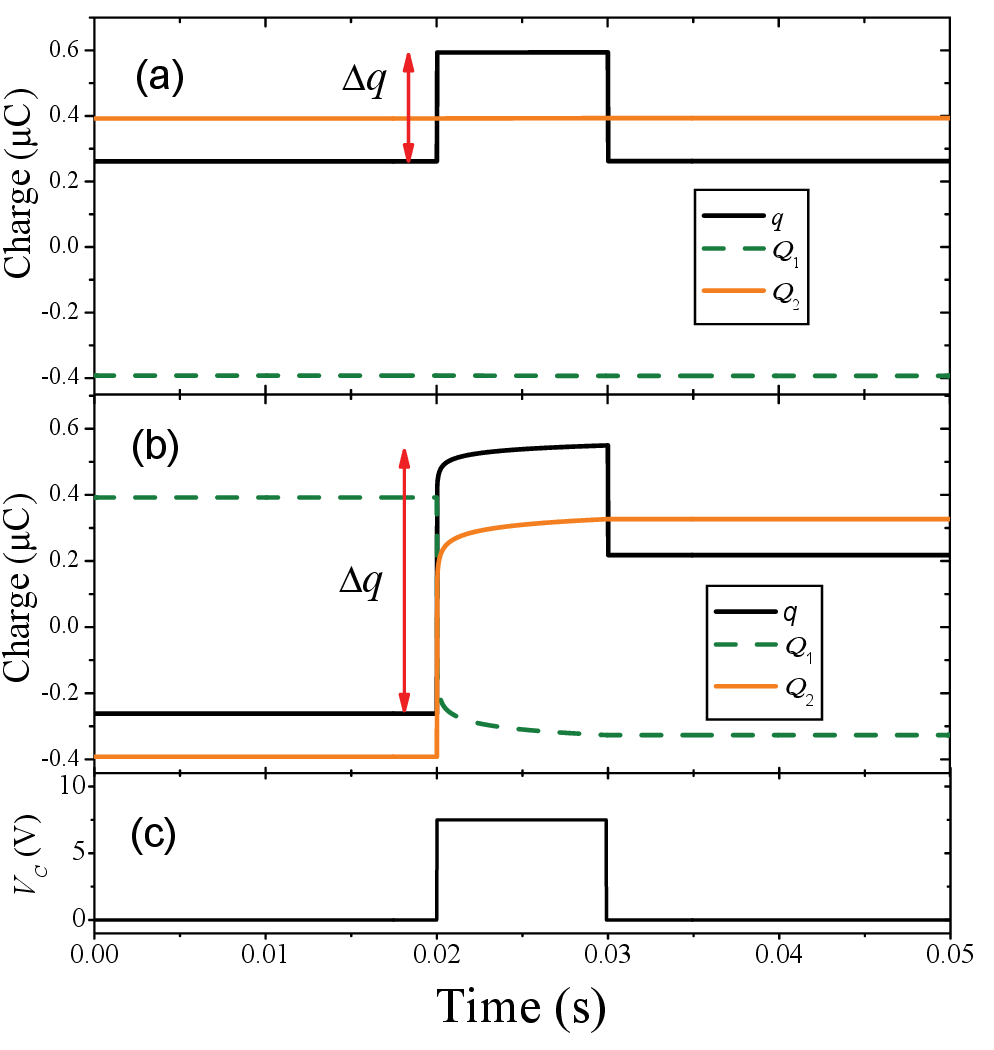}
\caption{The amount of charge $\Delta q$ needed for the same
change in $V_C$ (by 7.5V in the present calculation) depends on the
initial memcapacitor polarization. (a) and (b) represent the
device dynamics at different initial conditions, while (c) is the
voltage pulse profile. The memcapacitor parameters are as in Fig.
\ref{fig3}.} \label{fig6}
\end{figure}

Fig. \ref{fig6} illustrates the reading of the internal layer
polarization in a two-layer memcapacitor. Quite generally, we
start with a memcapacitor having a certain internal polarization
and assume that at the initial moment of time the voltage drop on
the memcapacitor is zero. Although this is not a steady state of
the device (at $t\rightarrow\infty$ and $V_C=0$ we expect
$q,Q_k\rightarrow 0$), the memcapacitor can stay in such a state
sufficiently long since the tunneling current is exponentially
suppressed at low internal fields. Our idea is to apply a single
voltage pulse to the device and monitor the amount of charge
needed to make $V_C$ close to the applied voltage $V$. This amount
of charge $\Delta q$, as we show, depends on the device state.

In our particular calculations, we consider two cases
characterized by a different initial polarization. When $Q_1$ is
negative at $t=0$ (Fig. \ref{fig6}(a)), the applied voltage pulse
does not significantly change the device state ($Q_1$, $Q_2$) and
the amount of charge $\Delta q$ added to the plates is small
(hence the capacitance is low). In the opposite case, when $Q_1$
is positive at $t=0$ (Fig. \ref{fig6}(b)), much larger charge
should be transferred to the capacitor plates since
``re-programming'' of the internal state is required. Therefore,
this measurement would monitor a higher value of capacitance. We
emphasize that the measured capacitance value is determined by
both the system state and by the voltage probe. For the same
system state but different probes (for example, pulses of
different polarity but of same magnitude) the measured value of
capacitance may be different.

\section{Equivalent circuit model} \label{sec4}

In this Section, we suggest an equivalent circuit model of the
solid-state memcapacitor discussed above. This correspondence is very useful in circuit simulators, or in the realization of electronic circuits
to reproduce memcapacitive features. The main idea behind the
circuit model is to represent each insulator spacing between
adjacent metal layers (including capacitor plates) by a separate
capacitor and to use non-linear resistors in order to mimic the
electron tunneling between the layers and energy loss in the
thermalization processes accompanying the tunneling. The top and
bottom surfaces of each internal layer are considered as plates of
adjacent circuit model's capacitors while the metallic material of
the layer itself plays the role of the conductor connecting these
capacitors.

\begin{figure}[tb]
\centering
\includegraphics[width=6.5cm]{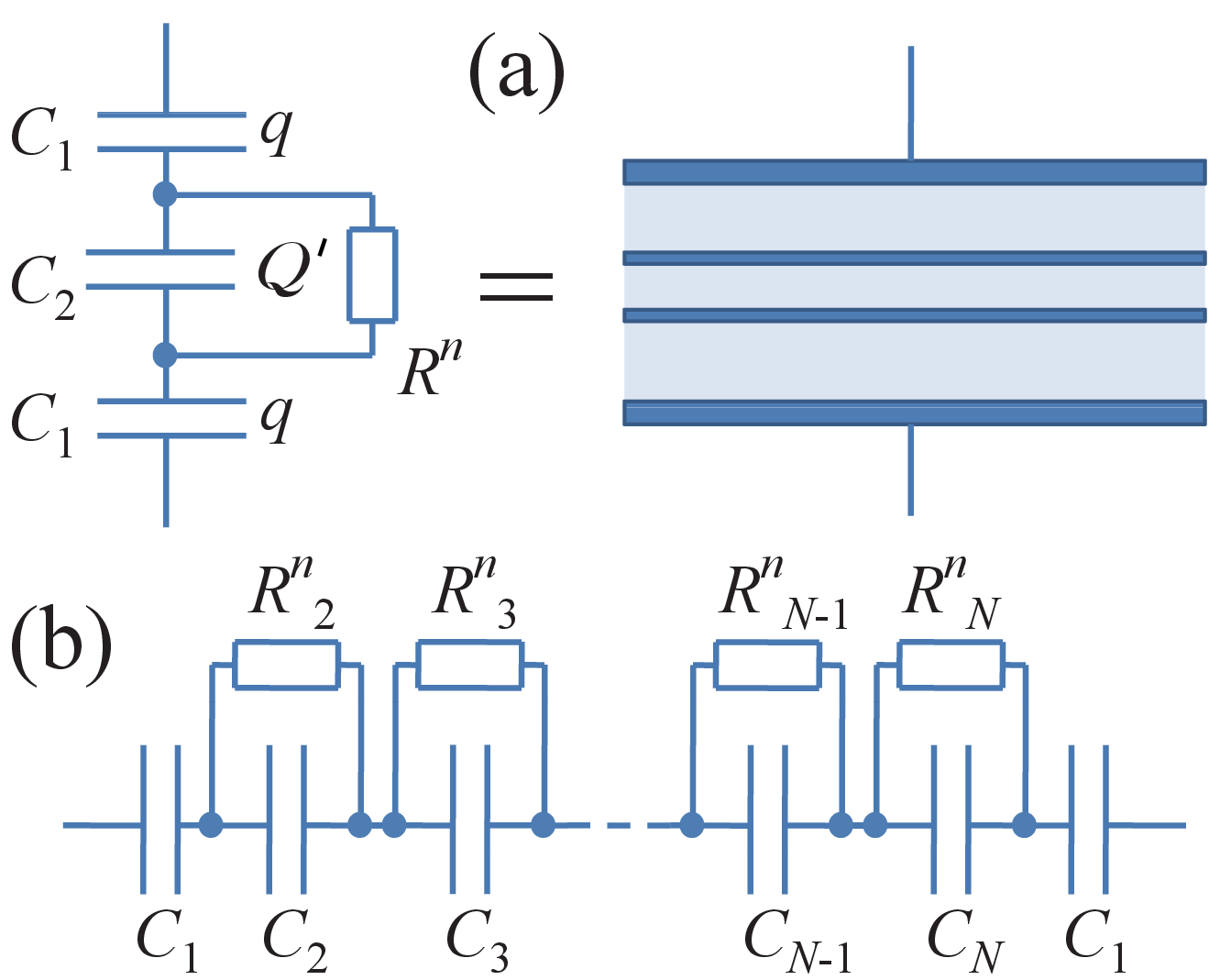}
\caption{(a) Equivalent circuit model of two-layer memcapacitor.
Here, $C_1$ and $C_2$ are usual capacitors and $R^{n}$ is a
non-linear resistor. It is assumed that the internal layers are
symmetrically embedded into the device. Otherwise, the capacitance values of
all capacitors would be different. (b) Equivalent circuit model of
$N$-layer memcapacitor.} \label{fig7}
\end{figure}

Fig. \ref{fig7}(a) shows the equivalent circuit of a two-layer
memcapacitor. The circuit is composed by three capacitors connected
in series. The capacitor in the middle is connected in parallel to
a non-linear resistor (see Fig. \ref{fig7}(a)). The parameters of
capacitors are determined by the usual parallel plate capacitor
expression and related to the parameters of memcapacitor in the
following way:
\begin{eqnarray}
C_1&=&\frac{\varepsilon_0\varepsilon_r\,S}{(d-\delta)/2}=\frac{2\,C_0}{1-\Delta},\nonumber\\
C_2&=&\frac{\varepsilon_0\varepsilon_r\,S}{\delta}=\frac{C_0}{\Delta}.\nonumber
\end{eqnarray}

Next, we demonstrate the equivalence of the equations describing the
circuit model and those of the memcapacitor. The circuit shown in
Fig. \ref{fig7}(a) is described by the following two equations
\begin{eqnarray}
V_C&=&\frac{2q}{C_1}+\frac{Q'}{C_2} \label{eq:circuit1} \\
\frac{dq}{dt}&=&\frac{dQ'}{dt}+I_{R}, \label{eq:circuit2}
\end{eqnarray}
where the first equation is for the total voltage drop and the
second equation is the Kirchhoff's law for the currents. Here, $I_R$ stands for the
current through the non-linear resistor which depends on the
voltage drop on $C_2$ (equal to $Q'/C_2$ with $Q'$ the charge on capacitor $C_2$).

Solving Eqs. (\ref{eq:circuit1},\ref{eq:circuit2}) for the total capacitance we obtain
\begin{eqnarray}
C=\frac{q}{V_C}=\frac{C_1\,C_2}{2C_2+C_1\frac{Q'}{q}}=\frac{C_0}{1+\Delta\frac{Q'-q}{q}},
\label{eq:circuit3}
\\ \frac{d\left(Q'-q\right)}{dt}=-I_{R}. \label{eq:circuit4}
\end{eqnarray}

In order to relate Eqs. (\ref{eq:circuit3},\ref{eq:circuit4}) to the
memcapacitor equations from Sec. \ref{sec1}, we note that the sum
of charges at the bottom plate of $C_1$ (equal to $-q$) and at
the top plate of $C_2$ (equal to $Q'$, see Fig. \ref{fig7}(a)) is
equal to $Q_1$, that is the charge at the first layer of
memcapacitor, i.e., $Q_1=Q'-q$. One then immediately notices that Eq.
(\ref{eq:circuit3}) and Eq. (\ref{eqC1}) (for the case of two
layers) are exactly the same; Eq. (\ref{eq:circuit4}) and Eq.
(\ref{eq:charge_dynamics}) are also the same. This proofs the complete
equivalence between memcapacitor's and circuit model's equations.

The extension of the circuit model to an $N$-layer memcapacitor is
straightforward: $N-1$ different capacitors and non-linear
resistors must be introduced between the external capacitors
$C_1$. Fig. \ref{fig7}(b) shows the equivalent circuit model of an
$N$-layer memcapacitor. Another interesting model extension can be
achieved if a memristive dielectric material (e.g., VO$_2$ films~\cite{driscoll09b}) is used instead of
the usual dielectric in the device. Then, in the equivalent
circuit model, the role of non-linear resistors will be played by
memristors.

\section{Conclusion} \label{sec5}

In conclusion, we have suggested and analyzed a solid-state
memcapacitor made of a multi-layer structure embedded in a
capacitor. The resulting system has noteworthy features, such a
frequency-dependent hysteresis under an ac-voltage, diverging and
negative capacitance. These feature emerge due to the slow
polarizability of the internal multi-layer structure as a
consequence of electron tunneling between the internal metal
layers. This gives rise to both complete screening of the field
due to the external metal plates - giving rise to diverging
capacitances - and over-screening of the same field leading to
negative capacitances. Clearly, when combined to an external
circuit with finite capacitance, the capacitance divergencies of
our proposed solid-state capacitor would be cut off by the
external circuit. Nonetheless, both the information stored in this
memcapacitor in a continuous fashion (analog operation) and its
negative capacitance in a certain range of the external field, may
find useful applications in electronics. In particular, we have
suggested a simple way to read the information stored in the
capacitor by using appropriate single pulses.

We have also shown the equivalent circuit model for this
memcapacitor based on a combination of regular capacitors and
non-linear resistors. This analogy could be useful both in actual
calculations with circuit simulators, or in experiments with
electronic circuits to reproduce memcapacitive features. Finally,
it is worth stressing that the memcapacitor presented in this work
can be easily fabricated by growing a metamaterial of several
metallic layers separated by an insulator (e.g., an oxide) in
between the metallic plates of a regular capacitor. Since only
``standard'' materials (as opposed to ferroelectrics) are used,
the suggested memcapacitor should also show high reliability.
Several modifications of the suggested structure are possible,
such as suggested in the text use of phase change or memristive
materials as well as e.g. use of nano-size metal beads instead of
metal layers. We thus hope this work will motive experimental
research in this direction.

\section*{Acknowledgment}
We thank D.N. Basov for useful discussions. This work has been
partially funded by the NSF grant No. DMR-0802830.

\bibliography{memcapacitor}

\begin{thebibliography}{22}
\expandafter\ifx\csname natexlab\endcsname\relax\def\natexlab#1{#1}\fi
\expandafter\ifx\csname bibnamefont\endcsname\relax
  \def\bibnamefont#1{#1}\fi
\expandafter\ifx\csname bibfnamefont\endcsname\relax
  \def\bibfnamefont#1{#1}\fi
\expandafter\ifx\csname citenamefont\endcsname\relax
  \def\citenamefont#1{#1}\fi
\expandafter\ifx\csname url\endcsname\relax
  \def\url#1{\texttt{#1}}\fi
\expandafter\ifx\csname urlprefix\endcsname\relax\def\urlprefix{URL }\fi
\providecommand{\bibinfo}[2]{#2}
\providecommand{\eprint}[2][]{\url{#2}}

\bibitem[{\citenamefont{Strukov et~al.}(2008)\citenamefont{Strukov, Snider,
  Stewart, and Williams}}]{strukov08}
\bibinfo{author}{\bibfnamefont{D.~B.} \bibnamefont{Strukov}},
  \bibinfo{author}{\bibfnamefont{G.~S.} \bibnamefont{Snider}},
  \bibinfo{author}{\bibfnamefont{D.~R.} \bibnamefont{Stewart}},
  \bibnamefont{and} \bibinfo{author}{\bibfnamefont{R.~S.}
  \bibnamefont{Williams}}, \bibinfo{journal}{Nature}
  \textbf{\bibinfo{volume}{453}}, \bibinfo{pages}{80} (\bibinfo{year}{2008}).

\bibitem[{\citenamefont{Pershin and {Di Ventra}}(2008)}]{pershin08}
\bibinfo{author}{\bibfnamefont{Y.~V.} \bibnamefont{Pershin}} \bibnamefont{and}
  \bibinfo{author}{\bibfnamefont{M.}~\bibnamefont{{Di Ventra}}},
  \bibinfo{journal}{Phys. Rev. B} \textbf{\bibinfo{volume}{78}},
  \bibinfo{pages}{113309} (\bibinfo{year}{2008}).

\bibitem[{\citenamefont{Pershin et~al.}(2009)\citenamefont{Pershin, {La
  Fontaine}, and {Di Ventra}}}]{pershin09}
\bibinfo{author}{\bibfnamefont{Y.~V.} \bibnamefont{Pershin}},
  \bibinfo{author}{\bibfnamefont{S.}~\bibnamefont{{La Fontaine}}},
  \bibnamefont{and} \bibinfo{author}{\bibfnamefont{M.}~\bibnamefont{{Di
  Ventra}}}, \bibinfo{journal}{Phys. Rev. E} \textbf{\bibinfo{volume}{80}},
  \bibinfo{pages}{021926} (\bibinfo{year}{2009}).

\bibitem[{\citenamefont{Driscoll
  et~al.}(2009{\natexlab{a}})\citenamefont{Driscoll, Kim, Chae, Kim, Lee,
  Jokerst, Palit, Smith, {Di Ventra}, and Basov}}]{driscoll09a}
\bibinfo{author}{\bibfnamefont{T.}~\bibnamefont{Driscoll}},
  \bibinfo{author}{\bibfnamefont{H.-T.} \bibnamefont{Kim}},
  \bibinfo{author}{\bibfnamefont{B.-G.} \bibnamefont{Chae}},
  \bibinfo{author}{\bibfnamefont{B.-J.} \bibnamefont{Kim}},
  \bibinfo{author}{\bibfnamefont{Y.-W.} \bibnamefont{Lee}},
  \bibinfo{author}{\bibfnamefont{N.~M.} \bibnamefont{Jokerst}},
  \bibinfo{author}{\bibfnamefont{S.}~\bibnamefont{Palit}},
  \bibinfo{author}{\bibfnamefont{D.~R.} \bibnamefont{Smith}},
  \bibinfo{author}{\bibfnamefont{M.}~\bibnamefont{{Di Ventra}}},
  \bibnamefont{and} \bibinfo{author}{\bibfnamefont{D.~N.} \bibnamefont{Basov}},
  \bibinfo{journal}{Science} \textbf{\bibinfo{volume}{325}},
  \bibinfo{pages}{1518} (\bibinfo{year}{2009}{\natexlab{a}}).

\bibitem[{\citenamefont{Driscoll
  et~al.}(2009{\natexlab{b}})\citenamefont{Driscoll, Kim, Chae, {Di Ventra},
  and Basov}}]{driscoll09b}
\bibinfo{author}{\bibfnamefont{T.}~\bibnamefont{Driscoll}},
  \bibinfo{author}{\bibfnamefont{H.-T.} \bibnamefont{Kim}},
  \bibinfo{author}{\bibfnamefont{B.~G.} \bibnamefont{Chae}},
  \bibinfo{author}{\bibfnamefont{M.}~\bibnamefont{{Di Ventra}}},
  \bibnamefont{and} \bibinfo{author}{\bibfnamefont{D.~N.} \bibnamefont{Basov}},
  \bibinfo{journal}{Appl. Phys. Lett.} \textbf{\bibinfo{volume}{95}},
  \bibinfo{pages}{043503} (\bibinfo{year}{2009}{\natexlab{b}}).

\bibitem[{\citenamefont{Waser and Aono}(2007)}]{Waser2007-1}
\bibinfo{author}{\bibfnamefont{R.}~\bibnamefont{Waser}} \bibnamefont{and}
  \bibinfo{author}{\bibfnamefont{M.}~\bibnamefont{Aono}},
  \bibinfo{journal}{Nature {M}aterials} \textbf{\bibinfo{volume}{6}},
  \bibinfo{pages}{833} (\bibinfo{year}{2007}).

\bibitem[{\citenamefont{Scott and Bozano}(2007)}]{Scott2007-1}
\bibinfo{author}{\bibfnamefont{J.~C.} \bibnamefont{Scott}} \bibnamefont{and}
  \bibinfo{author}{\bibfnamefont{L.~D.} \bibnamefont{Bozano}},
  \bibinfo{journal}{Advanced {M}aterials} \textbf{\bibinfo{volume}{19}},
  \bibinfo{pages}{1452} (\bibinfo{year}{2007}).

\bibitem[{\citenamefont{{Di Ventra}
  et~al.}(2009{\natexlab{a}})\citenamefont{{Di Ventra}, Pershin, and
  Chua}}]{diventra09}
\bibinfo{author}{\bibfnamefont{M.}~\bibnamefont{{Di Ventra}}},
  \bibinfo{author}{\bibfnamefont{Y.~V.} \bibnamefont{Pershin}},
  \bibnamefont{and} \bibinfo{author}{\bibfnamefont{L.~O.} \bibnamefont{Chua}},
  \bibinfo{journal}{Proc. IEEE} \textbf{\bibinfo{volume}{97}},
  \bibinfo{pages}{1717} (\bibinfo{year}{2009}{\natexlab{a}}).

\bibitem[{\citenamefont{{Di Ventra}
  et~al.}(2009{\natexlab{b}})\citenamefont{{Di Ventra}, Pershin, and
  Chua}}]{ourPointofView}
\bibinfo{author}{\bibfnamefont{M.}~\bibnamefont{{Di Ventra}}},
  \bibinfo{author}{\bibfnamefont{Y.~V.} \bibnamefont{Pershin}},
  \bibnamefont{and} \bibinfo{author}{\bibfnamefont{L.~O.} \bibnamefont{Chua}},
  \bibinfo{journal}{Proc. IEEE} \textbf{\bibinfo{volume}{97}},
  \bibinfo{pages}{1371} (\bibinfo{year}{2009}{\natexlab{b}}).

\bibitem[{\citenamefont{Kim et~al.}(2001)\citenamefont{Kim, Park, Chung, Bark,
  Yi, Choi, Kim, Lee, and Lee}}]{Kim2001-1}
\bibinfo{author}{\bibfnamefont{Y.}~\bibnamefont{Kim}},
  \bibinfo{author}{\bibfnamefont{K.~H.} \bibnamefont{Park}},
  \bibinfo{author}{\bibfnamefont{T.~H.} \bibnamefont{Chung}},
  \bibinfo{author}{\bibfnamefont{H.~J.} \bibnamefont{Bark}},
  \bibinfo{author}{\bibfnamefont{J.~Y.} \bibnamefont{Yi}},
  \bibinfo{author}{\bibfnamefont{W.~C.} \bibnamefont{Choi}},
  \bibinfo{author}{\bibfnamefont{E.~K.} \bibnamefont{Kim}},
  \bibinfo{author}{\bibfnamefont{J.~W.} \bibnamefont{Lee}}, \bibnamefont{and}
  \bibinfo{author}{\bibfnamefont{J.~Y.} \bibnamefont{Lee}},
  \bibinfo{journal}{{A}ppl. {P}hys. {L}ett.} \textbf{\bibinfo{volume}{78}},
  \bibinfo{pages}{934} (\bibinfo{year}{2001}).

\bibitem[{\citenamefont{Fleetwood et~al.}(1995)\citenamefont{Fleetwood,
  Shaneyfelt, Warren, Schwank, Meisenheimer, and Winokur}}]{fleetwood95}
\bibinfo{author}{\bibfnamefont{D.~M.} \bibnamefont{Fleetwood}},
  \bibinfo{author}{\bibfnamefont{M.~R.} \bibnamefont{Shaneyfelt}},
  \bibinfo{author}{\bibfnamefont{W.~L.} \bibnamefont{Warren}},
  \bibinfo{author}{\bibfnamefont{J.}~\bibnamefont{Schwank}},
  \bibinfo{author}{\bibfnamefont{T.~L.} \bibnamefont{Meisenheimer}},
  \bibnamefont{and} \bibinfo{author}{\bibfnamefont{P.~S.}
  \bibnamefont{Winokur}}, \bibinfo{journal}{Microelectronics and Reliability}
  \textbf{\bibinfo{volume}{35}}, \bibinfo{pages}{403} (\bibinfo{year}{1995}).

\bibitem[{\citenamefont{Su et~al.}(2005)\citenamefont{Su, Wang, Pilkuhn, and
  Pei}}]{su05}
\bibinfo{author}{\bibfnamefont{A.~Y.-K.} \bibnamefont{Su}},
  \bibinfo{author}{\bibfnamefont{H.~L.} \bibnamefont{Wang}},
  \bibinfo{author}{\bibfnamefont{M.~H.} \bibnamefont{Pilkuhn}},
  \bibnamefont{and} \bibinfo{author}{\bibfnamefont{Z.}~\bibnamefont{Pei}},
  \bibinfo{journal}{Appl. Phys. Lett.} \textbf{\bibinfo{volume}{86}},
  \bibinfo{pages}{062110} (\bibinfo{year}{2005}).

\bibitem[{\citenamefont{Lee et~al.}(2006)\citenamefont{Lee, Lu, Dai, Chan,
  Jelenkovic, and Tong}}]{lee06}
\bibinfo{author}{\bibfnamefont{P.~F.} \bibnamefont{Lee}},
  \bibinfo{author}{\bibfnamefont{X.~B.} \bibnamefont{Lu}},
  \bibinfo{author}{\bibfnamefont{J.~Y.} \bibnamefont{Dai}},
  \bibinfo{author}{\bibfnamefont{H.~L.~W.} \bibnamefont{Chan}},
  \bibinfo{author}{\bibfnamefont{E.}~\bibnamefont{Jelenkovic}},
  \bibnamefont{and} \bibinfo{author}{\bibfnamefont{K.~Y.} \bibnamefont{Tong}},
  \bibinfo{journal}{Nanotechnology} \textbf{\bibinfo{volume}{17}},
  \bibinfo{pages}{1202} (\bibinfo{year}{2006}).

\bibitem[{\citenamefont{Nieminen et~al.}(2002)\citenamefont{Nieminen, Ermolov,
  Nybergh, Silanto, and Ryhanen}}]{Nieminen2002-1}
\bibinfo{author}{\bibfnamefont{H.}~\bibnamefont{Nieminen}},
  \bibinfo{author}{\bibfnamefont{V.}~\bibnamefont{Ermolov}},
  \bibinfo{author}{\bibfnamefont{K.}~\bibnamefont{Nybergh}},
  \bibinfo{author}{\bibfnamefont{S.}~\bibnamefont{Silanto}}, \bibnamefont{and}
  \bibinfo{author}{\bibfnamefont{T.}~\bibnamefont{Ryhanen}},
  \bibinfo{journal}{{J}. {M}icromech. {M}icroeng.}
  \textbf{\bibinfo{volume}{12}}, \bibinfo{pages}{177} (\bibinfo{year}{2002}).

\bibitem[{\citenamefont{Partensky}(2002)}]{partensky2002-1}
\bibinfo{author}{\bibfnamefont{M.~B.} \bibnamefont{Partensky}},
  \bibinfo{journal}{arXiv:physics/0208048}  (\bibinfo{year}{2002}).

\bibitem[{\citenamefont{Pershin and Di~Ventra}(2009)}]{pershin2009-5}
\bibinfo{author}{\bibfnamefont{Y.~V.} \bibnamefont{Pershin}} \bibnamefont{and}
  \bibinfo{author}{\bibfnamefont{M.}~\bibnamefont{Di~Ventra}},
  \bibinfo{journal}{arXiv:0910.1583}  (\bibinfo{year}{2009}).

\bibitem[{\citenamefont{Krems et~al.}(2009)\citenamefont{Krems, Pershin, and
  Di~Ventra}}]{krems2009-2}
\bibinfo{author}{\bibfnamefont{M.}~\bibnamefont{Krems}},
  \bibinfo{author}{\bibfnamefont{Y.~V.} \bibnamefont{Pershin}},
  \bibnamefont{and}
  \bibinfo{author}{\bibfnamefont{M.}~\bibnamefont{Di~Ventra}},
  \bibinfo{journal}{to be published}  (\bibinfo{year}{2009}).

\bibitem[{\citenamefont{Ershov et~al.}(1998)\citenamefont{Ershov, Liu, Li,
  Buchanan, Wasilewski, and Jonscher}}]{Ershov1998-1}
\bibinfo{author}{\bibfnamefont{M.}~\bibnamefont{Ershov}},
  \bibinfo{author}{\bibfnamefont{H.~C.} \bibnamefont{Liu}},
  \bibinfo{author}{\bibfnamefont{L.}~\bibnamefont{Li}},
  \bibinfo{author}{\bibfnamefont{M.}~\bibnamefont{Buchanan}},
  \bibinfo{author}{\bibfnamefont{Z.}~\bibnamefont{Wasilewski}},
  \bibnamefont{and} \bibinfo{author}{\bibfnamefont{A.~K.}
  \bibnamefont{Jonscher}}, \bibinfo{journal}{{IEEE} {T}ransactions on
  {E}lectron {D}evices} \textbf{\bibinfo{volume}{45}}, \bibinfo{pages}{2196}
  (\bibinfo{year}{1998}).

\bibitem[{\citenamefont{Gommans et~al.}(2005)\citenamefont{Gommans, Kemerink,
  and Janssen}}]{Gommans2005-1}
\bibinfo{author}{\bibfnamefont{H.~H.~P.} \bibnamefont{Gommans}},
  \bibinfo{author}{\bibfnamefont{M.}~\bibnamefont{Kemerink}}, \bibnamefont{and}
  \bibinfo{author}{\bibfnamefont{R.~A.~J.} \bibnamefont{Janssen}},
  \bibinfo{journal}{Phys. Rev. B} \textbf{\bibinfo{volume}{72}},
  \bibinfo{pages}{235204} (\bibinfo{year}{2005}).

\bibitem[{\citenamefont{Mora-Sero et~al.}(2006)\citenamefont{Mora-Sero,
  Bisquert, Fabregat-Santiago, Garcia-Belmonte, Zoppi, Durose, Proskuryakov,
  Oja, Belaidi, Dittrich et~al.}}]{Mora2006-1}
\bibinfo{author}{\bibfnamefont{I.}~\bibnamefont{Mora-Sero}},
  \bibinfo{author}{\bibfnamefont{J.}~\bibnamefont{Bisquert}},
  \bibinfo{author}{\bibfnamefont{F.}~\bibnamefont{Fabregat-Santiago}},
  \bibinfo{author}{\bibfnamefont{G.}~\bibnamefont{Garcia-Belmonte}},
  \bibinfo{author}{\bibfnamefont{G.}~\bibnamefont{Zoppi}},
  \bibinfo{author}{\bibfnamefont{K.}~\bibnamefont{Durose}},
  \bibinfo{author}{\bibfnamefont{Y.}~\bibnamefont{Proskuryakov}},
  \bibinfo{author}{\bibfnamefont{I.}~\bibnamefont{Oja}},
  \bibinfo{author}{\bibfnamefont{A.}~\bibnamefont{Belaidi}},
  \bibinfo{author}{\bibfnamefont{T.}~\bibnamefont{Dittrich}},
  \bibnamefont{et~al.}, \bibinfo{journal}{{N}ano {L}etters}
  \textbf{\bibinfo{volume}{6}}, \bibinfo{pages}{640} (\bibinfo{year}{2006}).

\bibitem[{\citenamefont{Simmons}({1963})}]{Simmons1963-1}
\bibinfo{author}{\bibfnamefont{J.~G.} \bibnamefont{Simmons}},
  \bibinfo{journal}{{J. Appl. Phys.}} \textbf{\bibinfo{volume}{{34}}},
  \bibinfo{pages}{1793} (\bibinfo{year}{{1963}}).

\bibitem[{\citenamefont{{Di Ventra}}(2008)}]{diventra08}
\bibinfo{author}{\bibfnamefont{M.}~\bibnamefont{{Di Ventra}}},
  \emph{\bibinfo{title}{Electrical Transport in Nanoscale Systems}}
  (\bibinfo{publisher}{Cambridge University Press}, \bibinfo{year}{2008}).

\end{thebibliography}
\end{document}